\def\cm{{\rm\thinspace cm}}

\def\erg{{\rm\thinspace erg}}

\def\km{{\rm\thinspace km}}
\def\kpc{{\rm\thinspace kpc}}

\def\Msun{\hbox{$\rm\thinspace M_{\odot}$}}

\def\s{{\rm\thinspace s}}
\def\yr{{\rm\thinspace yr}}

\def\ergps{\hbox{$\erg\s^{-1}\,$}}

\def\kmps{\hbox{$\km\s^{-1}\,$}}

\def\Msunpyr{\hbox{$\Msun\yr^{-1}\,$}}

\def\psqcm{\hbox{$\cm^{-2}\,$}}

\documentstyle[psfig,times]{mn}
\begin{document}
\title{Chandra imaging of the complex X-ray core of the Perseus cluster}
\author[]
{\parbox[]{6.in} {A.C.~Fabian, J.S.~Sanders, S.~Ettori, G.B.~Taylor$^*$,
S.W.~Allen, C.S.~Crawford, K.~Iwasawa, R.M.~Johnstone and
P.M.~Ogle$^{\dagger}$ \\
\footnotesize
Institute of Astronomy, Madingley Road, Cambridge CB3 0HA \\
$*$ National Radio Astronomy Observatory, P.O. Box 0, Socorro, NM
87801, USA\\
$\dagger$ Center for Space Research, Massachusetts Institute of
Technology, Cambridge MA 02139, USA}}

\maketitle
\begin{abstract}
We report subarcsec-resolution X-ray imaging of the core of the
Perseus cluster around the galaxy NGC\,1275 with the Chandra X-ray
Observatory. The ROSAT-discovered holes associated with the radio
lobes have X-ray bright rims which are cooler than the surrounding gas
and not due to shocks. The holes themselves may contain some hotter
gas. We map strong photoelectric absorption across the Northern lobe
and rim due to a small infalling irregular galaxy, known as the high
velocity system. Two outer holes, one of which was previously known,
are identified with recently found spurs of low-frequency radio
emission. The spiral appearance of the X-ray cooler gas and the outer
optical parts of NGC\,1275 may be due to angular momentum in the
cooling flow.
\end{abstract}

\begin{keywords}
galaxies: individual: Perseus -- cooling flows -- galaxies:
individual: NGC\,1275 -- X-rays: galaxies
\end{keywords}

\section{Introduction}

The Perseus cluster, Abell\,426, at a redshift $z=0.0183$ or distance
about 100~Mpc is the brightest cluster in the sky in X-rays. It hosts
the nearest large cooling flow (e.g. Fabian et al 1981; Allen et al
1990; Fabian et al 1994). X-ray analysis of ASCA spectra indicates
that the mass deposition rate is about $300\Msunpyr$ (Allen et al
1999). At the centre is the galaxy NGC\,1275, surrounded by a
spectacular low-ionization, emission line nebula (Lynds 1970). The
nucleus powers the radio source 3C84 (Pedlar et al 1990). ROSAT HRI
images of the central region around NGC1275 show two 'holes' in the
X-ray emission coincident with the 0.5 arcmin sized radio lobes of
3C84 (B\"ohringer et al 1993). The overlap of the holes with both
radio and optical images of NGC1275 have been discussed further by
McNamara, O'Connell
\& Sarazin (1996) and the overall X-ray structure by Churazov et al
(2000). Here we present {\it Chandra} observations of the core of the
Perseus cluster using the ACIS-S detector, covering the energy range
0.5--7~keV.

\section{The X-ray image}

\begin{figure}
\centerline{\psfig{figure=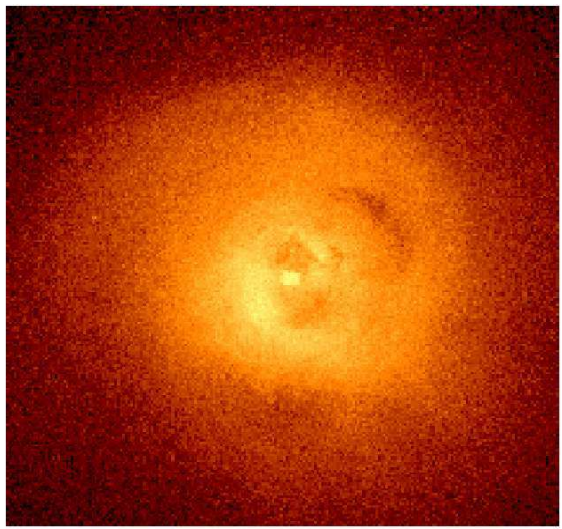,width=0.5\textwidth}}
\caption{Chandra image of the central, 6.5 arcmin wide, field of the Perseus
cluster in the 0.5--7~keV energy band. The pixel size is 1~arcsec,
North is to the top and East to the left. }
\end{figure}

The core of the Perseus cluster was observed with the Chandra X-ray
Observatory (Weisskopf et al 2000) on 2000 Jan 29 for 24,453~s. The
0.5--7~keV band image of the innermost 6.5 arcmin (195~kpc) diameter
region centred on NGC1275 is shown in Fig.~1. It has been exposure-map
corrected, chiefly to remove the effects of nodal readout boundary
structures. These mainly consist of 2 rows of pixels running from
close to the nucleus out to the west (p.a. 260 deg).

The holes in the emission associated with the inner radio lobes of
3C84 are clearly seen, as well as the larger hole to the NW, known
from Einstein Observatory images (Fabian et al 1981;
Branduardi-Raymont et al 1981). The inner radio holes in the
0.5--7~keV image are deep with a central surface brightness (20--25 ct
arcsec$^{-2}$ for the S hole and 15-30 ct arcsec$^{-2}$ for the N
hole) comparable to that of regions generally at a twice their radius
(Fig.~2). The holes are surrounded by bright rims at about 60 to 30 ct
arcsec$^{-2}$ to the N and 40 to 30 in the S. The simplest
interpretation is that the rims of bright emission are shells
enclosing the holes. A bright patch of emission to the E, about 40 by
15 arcmin yields about 50--80 ct arcsec$^{-2}$. The outer NW hole is
at 7 ct arcsec$^{-2}$ and is surrounded by emission at about 23 to 17
ct arcsec$^{-2}$.

\begin{figure}
\centerline{\psfig{figure=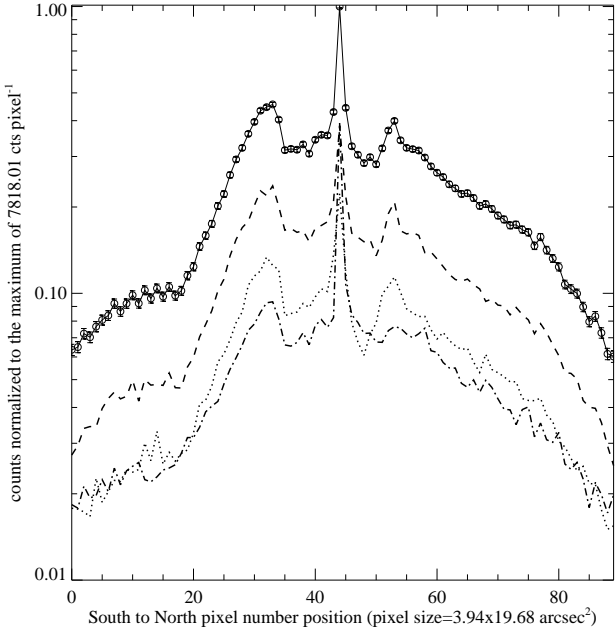,width=0.5\textwidth}}
\caption{Profile from S to N across the centre of the Perseus cluster.
The top curve is from the 0.5--7~keV band, with the 1--2, 0.5--1 and
2--7~keV profiles lying below (upper to lower).}
\end{figure}

\begin{figure}
\centerline{\psfig{figure=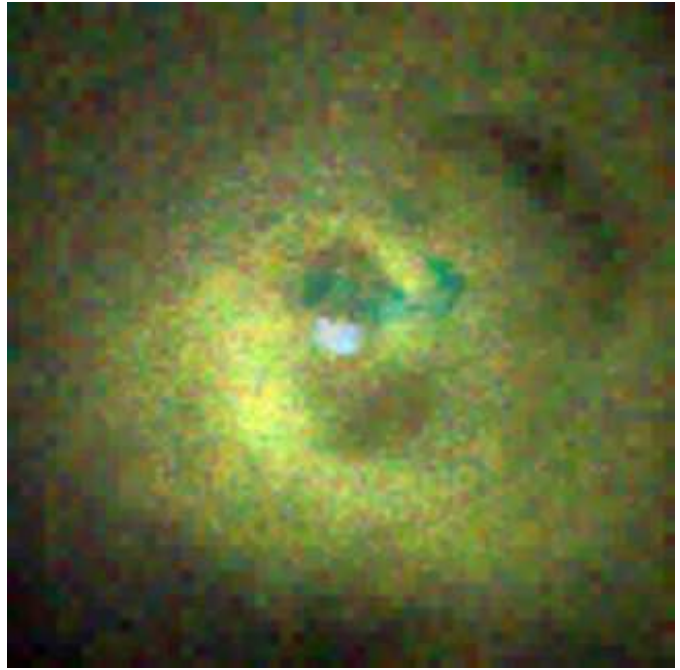,width=0.5\textwidth}}
\caption{X-ray colour image composed of exposure-corrected 0.5--1~keV
(red), 1--2~keV (green) and 2--7~keV (blue) images. The image is 3.5
arcmin square.}
\end{figure}

\begin{figure}
\centerline{\psfig{figure=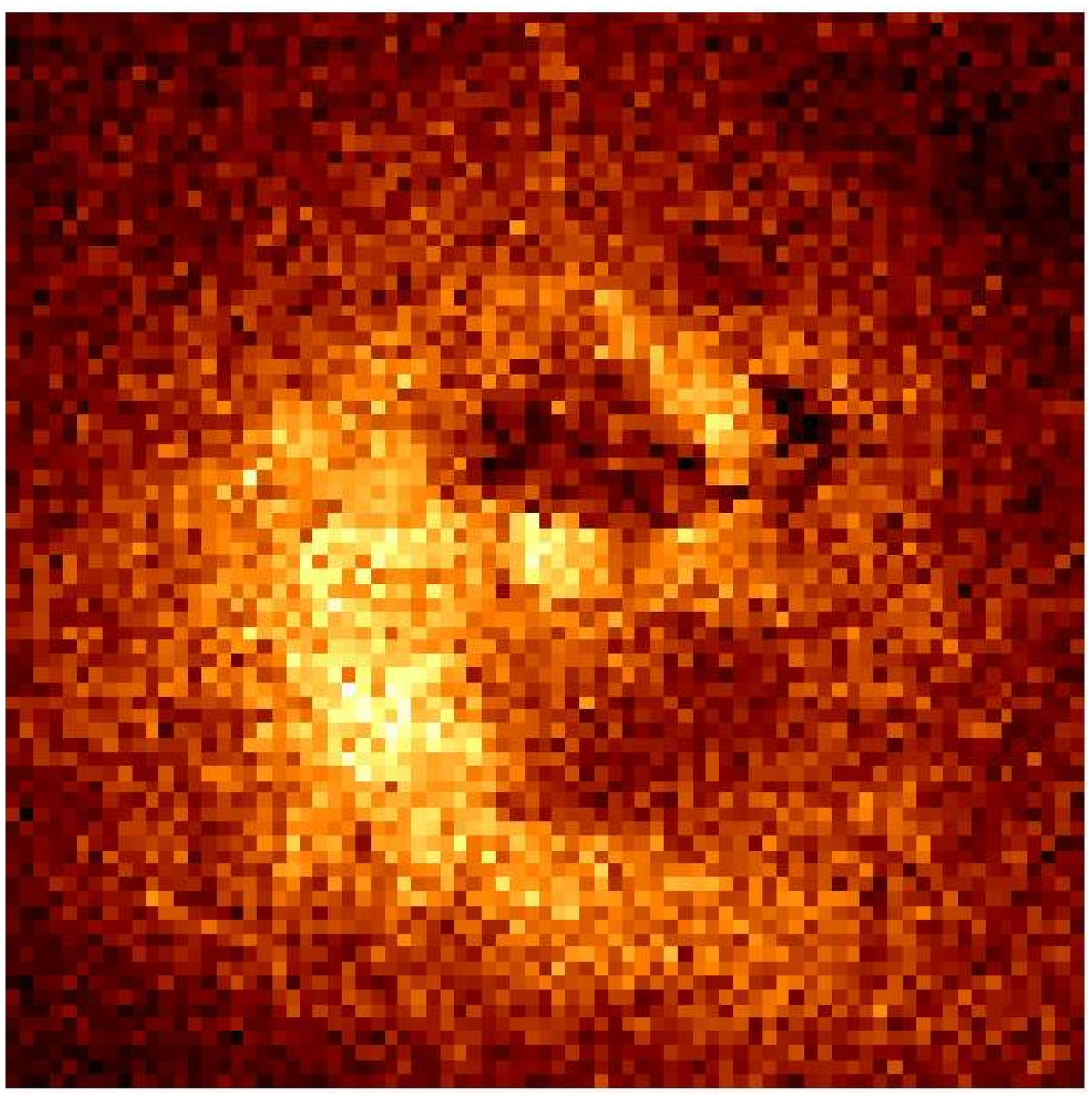,width=0.5\textwidth}}
\centerline{\psfig{figure=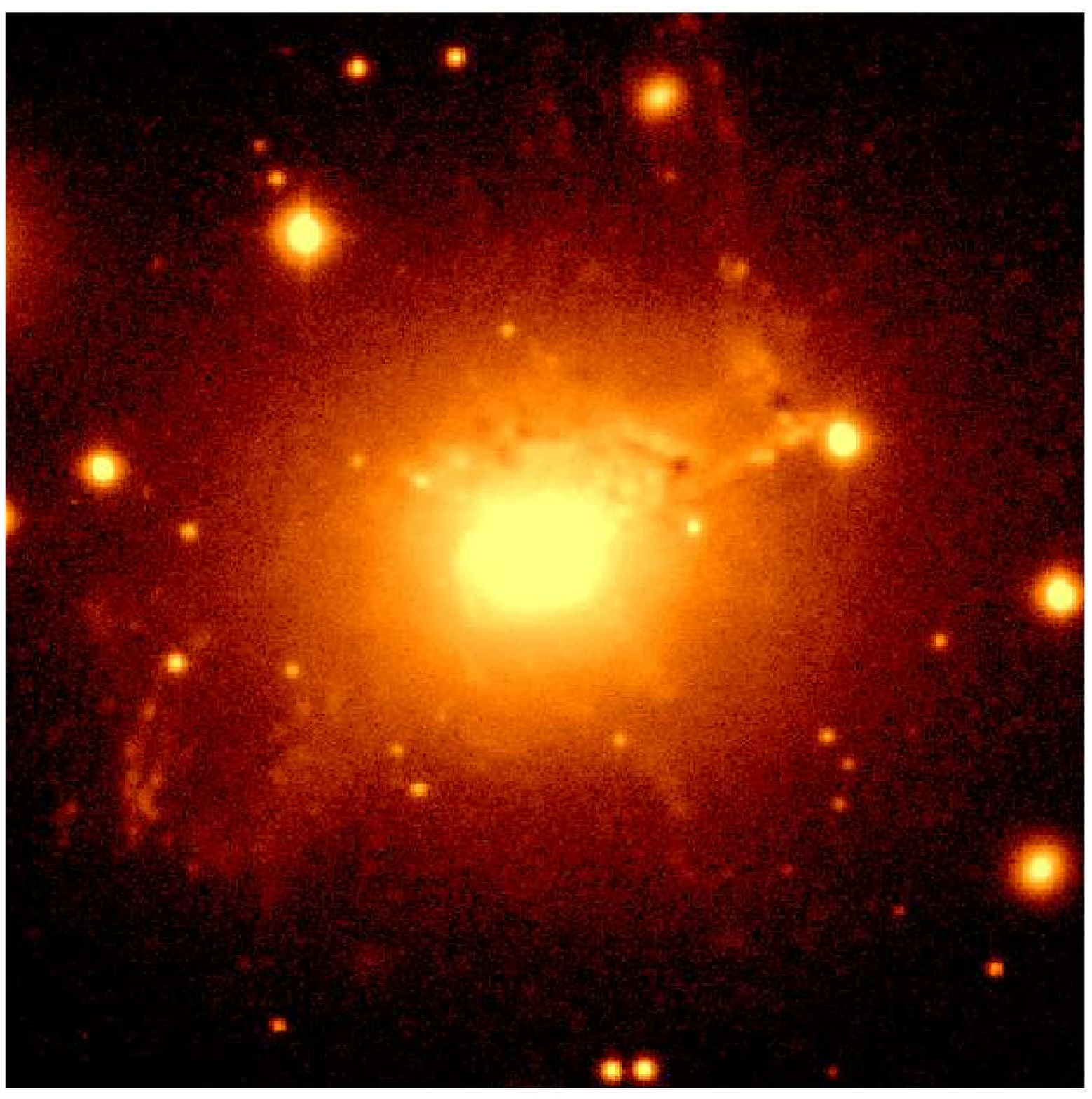,width=0.5\textwidth}}
\caption{Soft X-ray (0.5--1~keV) map (above) showing well the absorption
structure and a B-band optical image from the JKT (below) on the same
scale. The images are $150\times 150$ arcsec.}
\end{figure}

\begin{figure}
\centerline{\psfig{figure=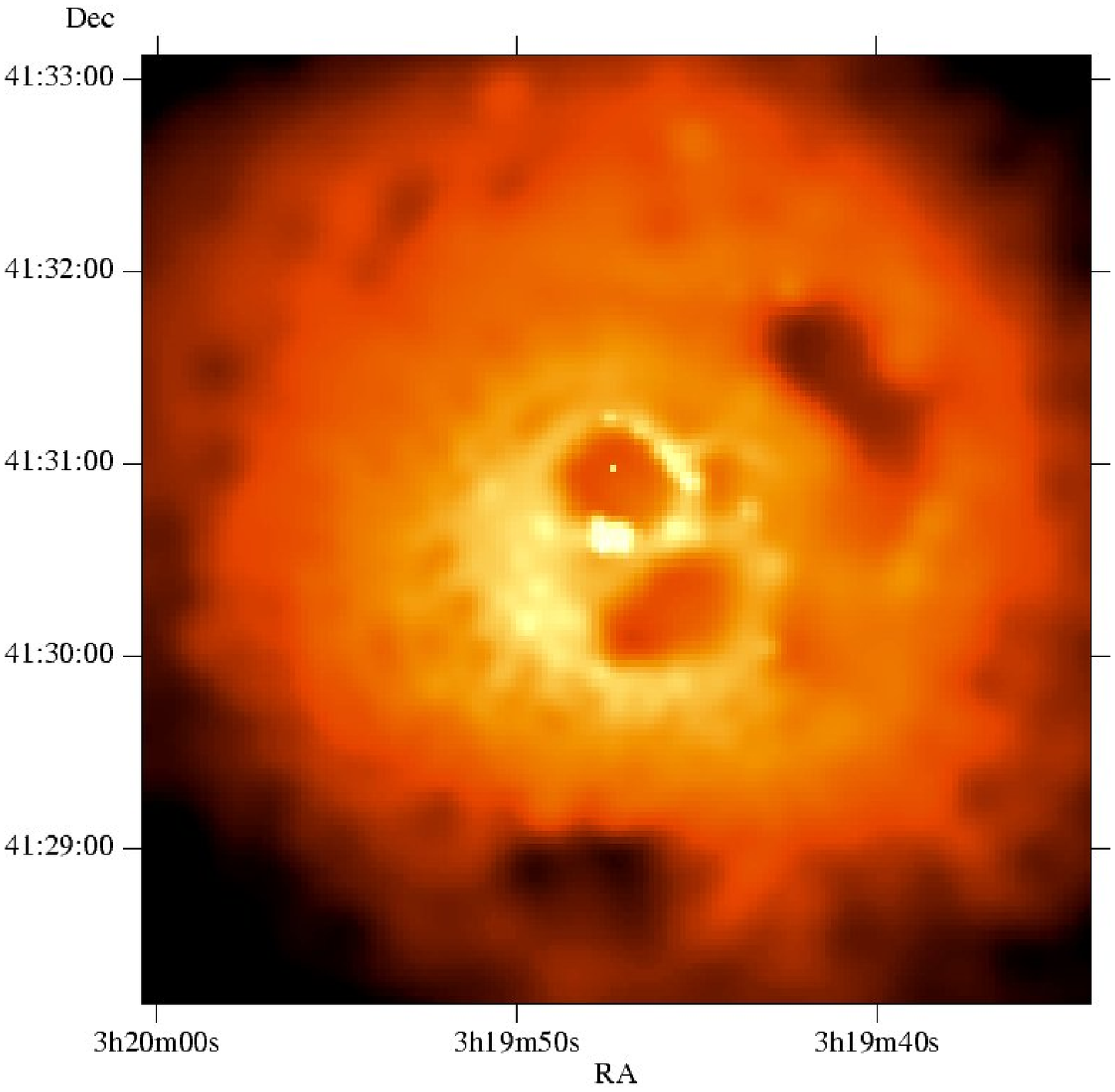,width=
0.4\textwidth}}
\centerline{\psfig{figure=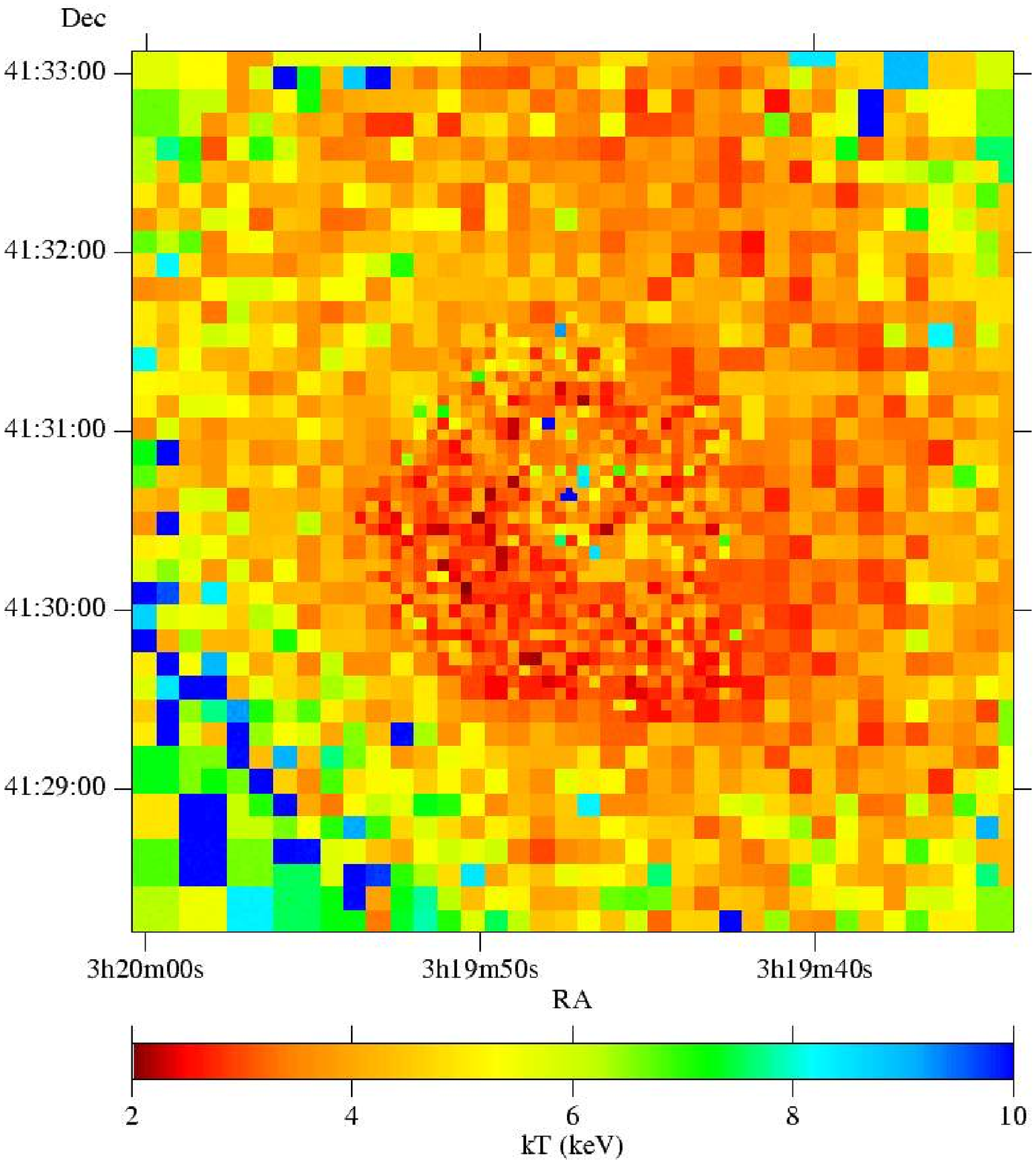,width=0.4\textwidth}}
\centerline{\psfig{figure=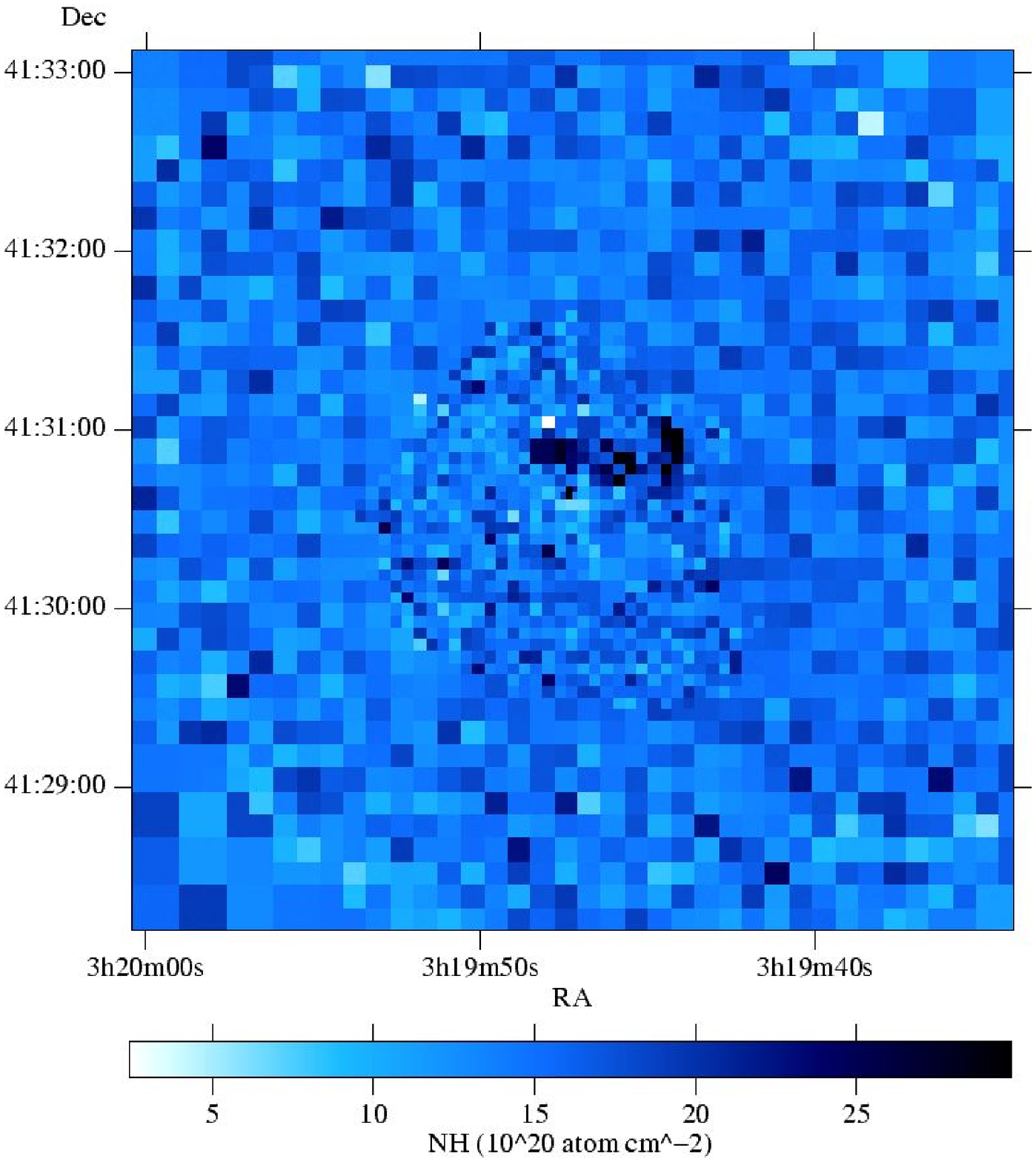,width=0.4\textwidth}}
\caption{Temperature (middle) and column density (bottom) maps from
adaptively binned images with an adaptively smoothed 0.5--7~keV image
(above). View the top image along pa 333 deg for a surprise.}
\end{figure}

\begin{figure}
\centerline{\psfig{figure=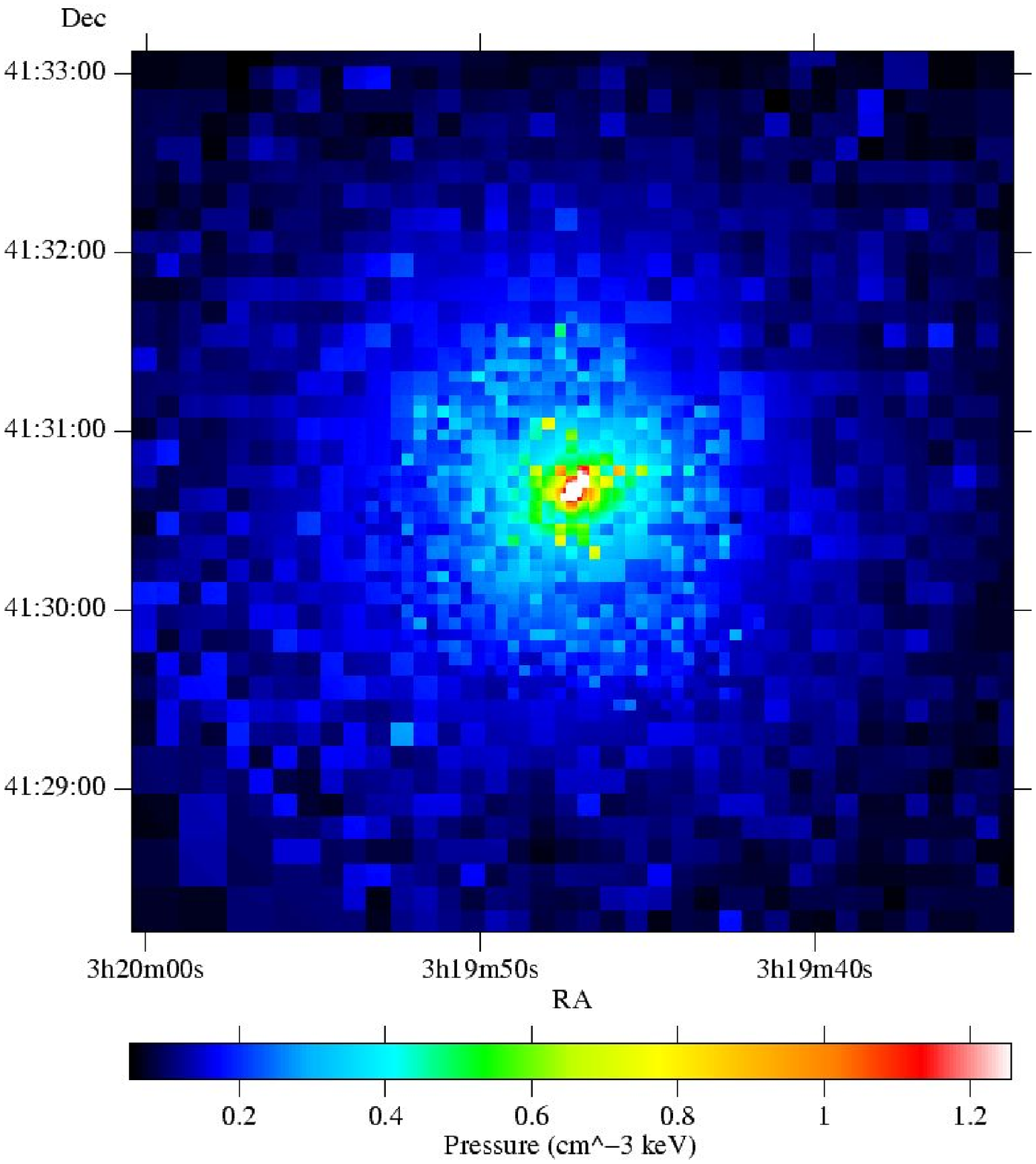,width=
0.4\textwidth}}
\centerline{\psfig{figure=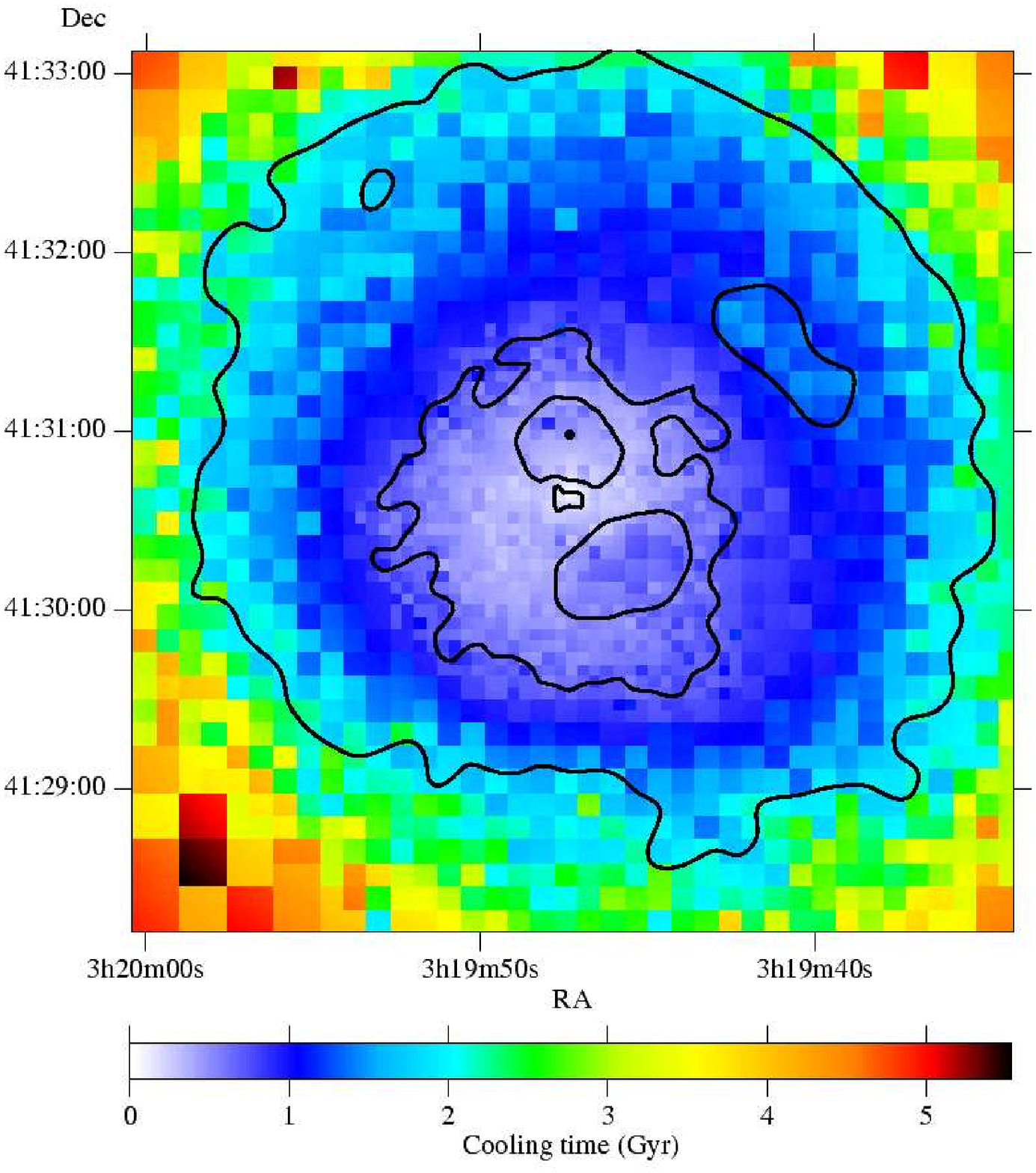,width=0.4\textwidth}}
\caption{Pressure (electron density times temperature; top) and
radiative cooling time (in Gyr from bremsstrahlung formula; bottom)
from the maps in Fig.~6.  }
\end{figure}

The nucleus of NGC\,1275 appears slightly extended and is spectrally
hard in our images. (Correction for the nodal readout boundary
structures, one of which passes close to the nucleus, has artificially
increased the apparent size of the nucleus in the images shown here.)
We defer discussion of its structure and detailed spectrum to future
work.

\section{X-ray colour analysis}

We have made soft (0.5--1~keV), medium (1--2~keV) and hard (2-7~keV)
X-ray maps in order to study the X-ray colours of the inner parts of
the cluster. They have been simply superimposed using the software
tool {\it GIMP} to provide the colour image shown in Fig.~3. It is
immediately seen that there are no strong colour gradients, apart from
some patches of hard emission (blue) across the N hole and its W rim.
The holes are clearly not due to absorption (or they would appear blue
in Fig.~3) and the rims are not hard but moderately soft, of the same
colours as the W bright patch. Inspection of separate colour images
bears this last result out; the rims are not distinguishable as sharp
features on a 3--7~keV image. They are therefore not shock features,
contrary to the prediction of Heinz, Reynolds \& Begelman (1998) and
the holes are not expanding supersonically.

The patchy absorption structure seen across the N hole and its W rim
is clearly seen in Fig.~4, which covers the 0.5--1~keV band. This
structure coincides in position and length with the high-velocity
emission-line nebulosity seen to the N of NGC\,1275 (Unger et al 1990
and references therein). This gas is thought to be associated with a
small irregular galaxy falling into the centre of the Perseus cluster
at a relative velocity of $3000\kmps$. Since it is seen in absorption
at 21~cm (de Young, Roberts \& Saslaw 1973) and Ly$\alpha$ (Briggs,
Snijders \& Boksenberg 1982), it must lie in front of NGC1275. Our
discovery of extended X-ray absorption by this galaxy means that it
must lie well in front of innermost regions, or such deep absorption
features would not be apparent. Without knowledge of the detailed
geometry of the holes and emission features we cannot establish any
precise estimate of its radial distance from the centre, but note that
the count rate at the bottom of the absorption patches is lower than
anywhere else within the inner 2 arcmin radius. This suggest that the
infalling galaxy may be more than 60~kpc from the cluster centre. The
westernmost part of the absorption structure appears to be surrounded
by brighter X-ray emission (Figs. 4 and 5). This could be shocked gas
within the infalling galaxy.

In order to quantify the temperature and absorption column-density
structure in the inner parts of the image we have compared colour
ratio images with tables of theoretical colour ratios. First we have
adaptively binned the data, after background subtraction, using a
local bin size which approximately equalizes the joint fractional
error per bin (over all colours) across the image. The counts from the
3 bands in each bin then provide the colour ratios. The tables were
predicted using the appropriate response matrices and auxilary
responses for the S3 detector and the MEKAL spectral code in XSPEC. A
metallicity of 0.4 times the cosmic value for all bins has been
assumed. Temperature and column density maps have been produced by
least-squares fitting of the observed colour ratios to the tables. An
adaptively smoothed (Ebeling et al 1999) image is included in Fig.~5
to illustrate the various features on the scale of the temperature and
column-density maps. The results confirm and quantify the appearance
obtained from inspection of the separate maps. The brighter emission
is due to gas which is at about 2.7~keV temperature, less than half
that of the outer gas which has a temperature of about 6.5~keV, in
good agreement with the ASCA results (e.g. Allen et al 2000). The
infalling galaxy has a mean excess column density of about
$10^{21}\psqcm$ which, over the region covered, corresponds to any
absorbing mass of about $2\times 10^9\Msun$ at cosmic abundance.

The maps have then been converted into (hot) electron density,
pressure and radiative cooling-time maps using the surface brightness
of the 1--2~keV map (the band with the highest count rate). The XSPEC
constructed tables enable the spectral normalization of the assumed
MEKAL models to be obtained and thereby the emission measure of each
pixel. We have then assumed that the path length through the emitting
gas is equal to the radius from the nucleus of that pixel, in order to
convert the emission measure into a density. This is a gross
assumption approximately correct for a symmetrical, highly-peaked
emission profile. Despite the holes etc in our image it should give a
rough indication of the gas properties (the density depends on the
inverse square root of the path length). We note that the pressure
inferred for the gas within a few kpc of the nucleus agrees with that
determined from the optical [SII] line ratio by Johnstone \& Fabian
(1988) and Heckman et al (1989). 

\section{Optical and radio}

The soft X-ray image is shown next to an optical B-band image in Fig.~4.
The B-band image was constructed from data obtained from the Isaac
Newton Group Archive. The image is the combination of images taken on
1997 Dec 25 and 1998 Aug 22 with the Jacobus Kapteyn Telescope (JKT) using
the Tek1024 detector, with a total exposure time of 17.7~ksec. The
data have been bias-subtracted and flat-fielded before being combined.
The correspondence between the soft X-ray absorption structure and the
high velocity system in the optical image is good. There are no other
obvious optical structures correlated with the X-ray image.

VLA Observations at a frequency of 1.4 GHz were combined from several
runs of 3--10 hours duration each taken between 1989 and 1999 in the
A, B and C configurations.  The resulting radio image, shown in
Fig. 7, has been restored with a 5 arcsec synthesised beam.  Contour
levels start at 1 mJy/beam and increase by factors of 2 to just below
the peak of 21.7 Jy/beam.

We note that the steepest part of the radio contours abut the
brightest part of the E patch.

\begin{figure}
\centerline{\psfig{figure=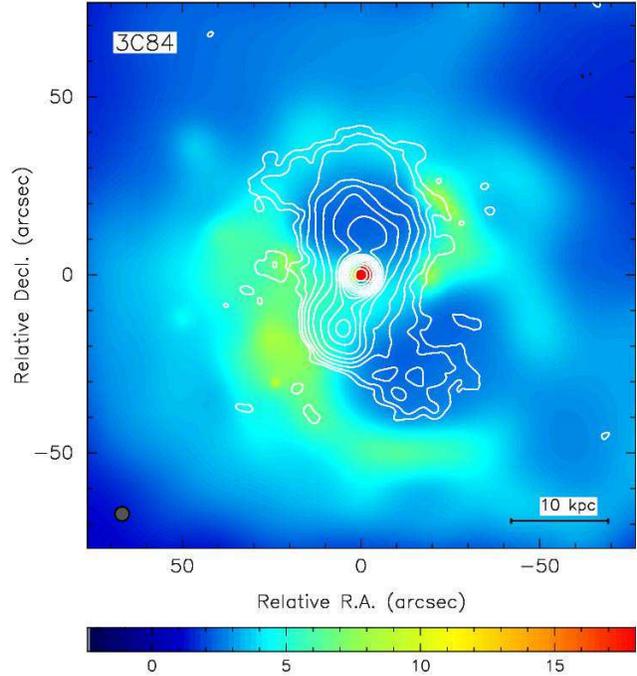,width=0.5\textwidth}}
\caption{Radio image overlay, see text for details. }
\end{figure}

\section{Discussion}

The overall X-ray appearance of the core of the Perseus cluster is of
a broadly circularly symmetric peak centred on NGC\,1275. The gas
temperature decreases inward with radius, from about 6.5 to less than
3~keV, as determined from the colour ratios assuming a single phase
gas. The radiative cooling time of the gas across the bright patch SE
of the nucleus is about $10^8\yr$. Therefore, unless there is a
balance with heating, the gas is part of a cooling flow (see Fabian
1994) and is expected to be multiphase, with the measured temperatures
being an emission-weighted mean. The radiative cooling time is less
than $6\times10^{9}\yr$ across the whole of the images shown.

The radio lobes of 3C84 occupy holes in the X-ray emission. The
simplest interpretation of the low surface brightness is that they are
devoid of X-ray gas and have pressure support from cosmic rays and
magnetic fields. However there may be hotter gas in the holes and
indeed we find that the pressure map shows little evidence for the
holes (the temperature map, Fig. 5, indicates 4--5~keV for the holes;
a result which is highly geometry dependent). The data are consistent
with the holes being filled with pressure-supporting hotter gas.
Future work involving detailed deprojection and spectral fits to the
data will enable this possibility to be carefully tested.

The composition of the radio lobes is important in determining the
energy content of the radio source and, using a synchrotron age, the
mean power output of the central engine. We have already noted that
there is no evidence for shocks so any expansion is subsonic. If the
holes only contain magnetic fields and cosmic rays then the energy is
approximately $P\Delta V\approx 8\times 10^{58}\erg$ (assuming that
the holes are spheres of radius 6~kpc each). A similar amount of work
must have been done in making the holes. That work probably causes a
very low frequency sound wave to propagate out from the hole region
and dissipate beyond the cluster core. There is no reason that it
should lead to local heating. The age of the inner holes is unclear
but probably of order $10^{7-8}\yr$, from synchroton ageing estimates
and also buoyancy considerations (Pedlar et al 1991; McNamara et al
2000; Churasov et al 2000). The mean power of the nucleus is then
$\sim 10^{44-45}\ergps$.

If the holes are just apparent and filled with hotter, lower surface
brightness gas than the rims, then it is difficult to estimate the
energy of the radio source. A probable lower estimate is the minimum
energy one of Pedlar et al (1990), which implies a pressure for the
relativistic fluid about ten times less than the local gas pressure.
The magnetic fields and/or cosmic rays need not be space-filling;
pressure balance is obtained if they occupy about 2 per cent of the
volume. The relativistic fluid may still have a large covering factor
and just pushes denser, cooler gas clouds out of the way, so that they
accumulate at the edge of the lobes, forming the bright rims. The
brightness of the rims relative to the surrounding region is
consistent with this hypothesis (i.e. $\Delta S/S\sim R/\Delta R$,
where $S$ and $R$ are surface brightness and radius, respectively).

The origin of the outer NW hole is unclear. It could, as suggested by
Churazov et al (2000), be a buoyant old radio lobe, or lobes, produced
by the central engine in NGC\,1275. If so then it must have
considerable surface tension, presumably due to ordered magnetic
fields, in order that it has not fallen apart. Perhaps the galaxy is
moving slowly to the SE and shed this lobe in the past. Such a
scenario could also account for the W bright patch as being where
cooler X-ray emitting clouds of gas are swept up by the radio lobes.

The situation is clarified somewhat by the recent low-frequency
(74~MHz) radio image of Blundell et al (2000) which shows a spur of
radio emission out to the position of the outer hole (there is also
some structure in the maps of Pedlar et al 1991 in this position). The
low-frequency map also shows a spur to the SSE, in the direction of
another hole seen in Fig.~5. Such low-frequency radio emission is
likely to originate from the oldest electron populations, thereby
supporting the idea that the spurs point to fossil radio lobes now
devoid of energetic electrons.

Of course, some of the apparent displacement of the outer holes from
the present jet (N--S) axis may be due to motion of the gas, rather
than the galaxy. If there is slow rotation of the gas then this could
explain the `swirly' appearance of the outer gas in Fig.~1 (noted by
Churazov et al 2000). It could be the remains an an earlier cooler
subcluster which merged with the cluster core. The impression of a
spiral structure winding inward in a clockwise sense is reinforced by
similar structures apparent in optical images (see e.g. Carlson et al
1998). We note that such an overall structure is consistent with a
cooling flow if the gas has some angular momentum which is roughly
conserved as the slow inflow proceeds. The streamlines of the flow
will be spiral rather than radial lines. These can then become
apparent if the gas at $\sim 50\kpc$ radius has some large-scale
inhomogeneity in the form of of wider spread of denser, cooler gas
over, say, one quadrant. As the gas cools the denser gas cools fastest
and as it flows inward creates an X-ray bright and cooler spiral. The
optical spiral is explained by star formation from the cooled gas.

Neither movement of gas nor galaxy obviously explains however the
outer S hole being along the present jet axis. The outer holes may
result as much from previous beaming directions of the central engine
as from its displacement. The situation resembles that around M87, as
found by B\"ohringer et al (1995) and most recently by Owen, Eilek \&
Kassim (1999). If this explanation is correct then it argues against
rapid spin of the central black hole being the origin of the jets. The
outer holes nevertheless do provide direct evidence that the central
radio source is sporadic and had earlier outbursts.

\section{Acknowledgements}
ACF is grateful to NASA for the opportunity to participate as an
InterDisciplinary Scientist and the Chandra project for such a superb
instrument. ACF, CSC and SWA thank the Royal Society for support.

\end{document}